\documentclass[conference,twocolumn,a4paper]{IEEEtran}

\usepackage{amsmath,amssymb}
\usepackage{graphicx,color}
\usepackage{url}
\usepackage{booktabs}

\def\SA{\mbox{SA}}

\def\proj{\mbox{\footnotesize proj}}
\def\env{\mbox{\footnotesize env}}
\DeclareMathOperator*{\argmin}{arg\,min}

\def\zZ{{\mathbb Z}}

\newtheorem{theorem}{Theorem}[section]

\newtheorem{fact}{Fact}[section]

\title{On Distributed Nonlinear Signal Analytics : Bandwidth and
Approximation Error Tradeoffs}

\author{\IEEEauthorblockN{Vijay Anavangot and Animesh
Kumar}
\IEEEauthorblockA{Department of Electrical Engineering \\
Indian Institute of Technology Bombay\\
Mumbai 400076, India \\
\{avijay,animesh\}@ee.iitb.ac.in}}

\begin{document}

\maketitle
\begin{abstract}
Analytics will be a part of the upcoming smart city and Internet of Things
(IoT). The focus of this work is approximate distributed \textit{signal
analytics}. It is envisaged that distributed IoT devices will record signals,
which may be of interest to the IoT cloud.  Communication of these signals from
IoT devices to the IoT cloud will require (lowpass) approximations.  Linear
signal approximations are well known in the literature. 
It will be outlined that in many IoT analytics problems, it is desirable that
the approximated signals (or their analytics) should always over-predict the
exact signals (or their analytics). This distributed
nonlinear approximation problem has not been studied before.  An algorithm
to perform distributed over-predictive signal analytics in the IoT cloud, based
on signal approximations by IoT devices, is proposed.  The \textit{fundamental
tradeoff} between the signal approximation bandwidth used by IoT devices and the
approximation error in signal analytics at the IoT cloud is quantified for the
class of differentiable signals.  Simulation results are also presented.
\end{abstract}

\section{Introduction}
\label{sec:intro}
Signal representation using a finite number of coefficients is well known
and is termed as source coding~\cite{cover2006},
sampling~\cite{shannon1949}, and function approximation~\cite{devore1993}. 
Analytics will be a part of the upcoming smart city and Internet
of Things (IoT). It is envisaged that distributed signals may be recorded at the
IoT devices~\cite{zanella2014}. At the IoT cloud, \textit{signal analytics} from
the recorded signals by many IoT~devices is of interest. 

Due to bandwidth constraint, each IoT~device should send signal approximation
for the intended signal analytics to the \textit{IoT cloud}. A parsimonious
signal approximation at each IoT~device, that minimizes approximation error at
the IoT cloud, is desired.  The fundamental tradeoff between approximation
bandwidth used by IoT devices and the approximation error in signal analytics at
the IoT cloud is desirable.

Linear or mean-squared signal approximations are well known in the literature.
An obvious method to approximate in a distributed manner is to compute the
signals' linear approximations using existing algorithms and communicate
them~\cite{devore1993,giridhar2005,vyavahare2016,jesus2015}.  However, this 
method
will not be applicable in certain nonlinear problems.  The following envisaged
smart city applications motivate distributed \textit{nonlinear} signal
analytics.
\begin{figure}[!htb]
\centering
\includegraphics[scale=0.82]{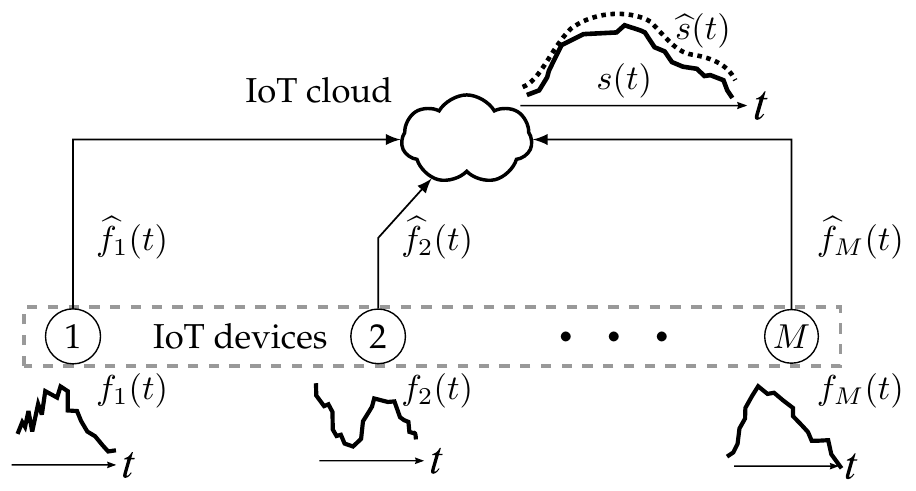}
\caption{\label{fig:SCApp} A schematic of nonlinear signal analytics in IoT is
  illustrated. IoT devices communicate their approximate signals $f_1(t),
  \ldots,f_M(t)$ to the IoT cloud in a distributed manner.}
\end{figure}

\emph{Smart meters}: Consider the setup shown in Fig.~\ref{fig:SCApp}. A smart
city planner fixes smart meters (IoT devices) in each home.  The smart meter
records the (instantaneous) power signal consumed, and has to report it to the
IoT cloud.  The planner has to calculate the smallest sufficient supply capacity
to meet the energy demand at all times.  If  $f_1(t), \ldots, f_M(t)$ are the
power consumption signals at various devices, then the planner is interested in
$\max_t\{f_1(t) + \ldots + f_M(t)\}$.  However, due to bandwidth constraint IoT
devices can only send approximations $\widehat{f}_i(t), \ldots,
\widehat{f}_M(t)$.  The IoT cloud will compute $\max_t\{\widehat{f}_1(t)+
\ldots+ \widehat{f}_M(t)\}$. For sufficient supply capacity calculation, it is
required that  
\begin{align}
\max_t\{\widehat{f}_1(t)+ \ldots+ \widehat{f}_M(t)\} \geq \max_t\{f_1(t) +
\ldots + f_M(t)\}. \nonumber
\end{align}
In the above equation, the difference between the two quantities is the
\textit{approximation error}. Using an (over-predictive) envelope approximation,
i.e., $\widehat{f}_i(t) \geq f_i(t), i = 1, \ldots, M$ is one approach to tackle
this problem. In this case, the IoT cloud reconstructs the sum
signal, $\widehat{s}(t) = \widehat{f}_1(t)+ \ldots + \widehat{f}_M(t)$ and then
reports $\max_t \widehat{s}(t)$ to the planner (see Fig.~\ref{fig:SCApp}). The
desired signal analytic and the associated approximation are both nonlinear.

\emph{Pollution control:} Consider a smart city regulator, which has to report
pollutant concentration (such as PM~2.5 levels). IoT devices with PM~2.5 sensors
can be employed in the smart city, which record $f_1(t),\ldots, f_M(t)$. To
save bandwidth, each IoT device has to communicate its approximated pollution
signal to the IoT cloud. The regulator can use air diffusion models
(see~\cite{Ranieri2012}) to characterize the spatio-temporal evolution of the
pollutant for the entire city. The regulator may want to provide a (pessimistic)
pollution signal approximation, which requires over-predictive nonlinear signal
analytics.

\emph{Renewable energy in a smart grid:} Consider a smart city
where solar panels generate electricity that returns to the power grid in the
region. Let $f_1(t), \ldots, f_M(t)$ be the renewable power generated as a
function of time $t$. These signals will be approximately communicated to the
IoT cloud. The electricity planner will be interested in $\min_{t}\{f_1(t)+
\ldots+ f_M(t)\}$ to estimate an (under-predictive) envelope of the renewable
power.  This is also a nonlinear signal analytics problem.

With a focus towards over-predictive nonlinear signal analytics in an IoT, the
following main results will be presented:
\begin{enumerate}
\item An algorithm using envelope approximations~\cite{kumar2017} is presented
for over-predictive signal analytics with IoT devices and cloud. Its simulation
results are presented.
\item For the above algorithm, with Fourier basis, fundamental tradeoffs between
approximation error and bandwidth will be analyzed for differentiable signals.
\end{enumerate}
\emph{Prior Art:} Linear signal analytics is well understood: distributed
signals can be projected (approximated) in a linear basis and approximately
communicated.  However, this method will not apply to the above-mentioned
applications. As far as we know, \emph{nonlinear signal analytics has not been
studied}.  On the other hand, analytics of distributed \textit{scalars} is well
reported.  Computation and reporting of symmetric functions of the scalar
parameters (such as weighted average) has been
studied~\cite{giridhar2005,vyavahare2016}. Gossip based algorithms compute
scalar functions in large networks,  using linear fusion
methods~\cite{jesus2015,tsitsiklis1984}.  Recently, there have been efforts
towards big-data analytics in smart 
city/IoT~\cite{zanella2014,sun2016,gubbi2013}. \textit{To the best of our
knowledge, these works do not address signal analytics where continuous-time
signals are involved.}

%

The paper is organized as follows. Section~\ref{sec:model} explains the system
background and problem formulation.  An order optimal approximation algorithm
for the over-predictive signal analytics is proposed in Section~\ref{sec:Algo}.
The bandwidth and error tradeoff for this scheme is analyzed in
Section~\ref{sec:MainRes}. Simulation setup and numerical results using
electricity load data sets are presented in Section~\ref{sec:Numsim}.
%
 
\section{Background and problem setup}
\label{sec:model}

IoT signals, their assumed smoothness properties, tradeoffs of
interest, and problem formulation are discussed in this section.

\subsection{Signal model and IoT description}
Consider an IoT with $M$ distributed IoT devices and an IoT cloud.  We
assume a \emph{cloud based IoT} architecture with computing capability at the
individual IoT devices \cite{gubbi2013}. The IoT cloud acts as a central server with the capacity
to handle massive data arriving from a number of IoT devices. The IoT
devices indexed $1, \ldots, M$ record signals $f_1(t), \ldots, f_M(t)$ at
various locations. Due to bandwidth constraint, the IoT devices individually
communicate approximations $\widehat{f}_1(t), \ldots, \widehat{f}_M(t)$ to the
IoT cloud. Signal analytics are derived in the IoT cloud based on
$\widehat{f}_1(t), \ldots, \widehat{f}_M(t)$. Without loss of generality, it is
assumed that $f_1(t), \ldots, f_M(t)$ are recorded over the finite observation
interval $t \in [0,1]$. The signals $f_1(t), \ldots, f_M(t)$ will be assumed to
be $p$-times differentiable in $t\in[0,1]$ for $p \geq 1$.

\subsection{Fourier representation of IoT signals}

Due to space constraints, this first exposition will use Fourier series basis
for $f_1(t), \ldots, f_M(t)$ on $[0,1]$, with the constraint $f_i(0) = f_i(1), i
= 1, \ldots, M$.\footnote{This end point symmetry constraint prevents Gibbs
phenomenon during reconstruction~\cite{oppenheim1999discrete} of the signals, 
and it
can be avoided by considering other basis (e.g., polynomials), but is omitted
due to space constraints~\cite{devore1993}.} The Fourier series of $f_1(t),
\ldots, f_M(t)$ are given by
\begin{align}
f_i(t) = \sum_{k=\infty}^{\infty} a_i[k] \exp(j 2 \pi k t),  \quad t \in [0,1],
i = 1, \ldots, M. \nonumber
\end{align}
The $p$-times differentiability of $f_i(t), i = 1, \ldots, M$ implies that their
Fourier coefficients decay polynomially in $k$:

\begin{fact}[Sec~2.3,\cite{mallat2008}]
\label{lem:regularity} 
A signal $f(t), t \in [0,1]$, with $f(0) = f(1)$, is $p$-times differentiable
if its Fourier coefficient~$a[k]$ obey
 \begin{align}
|a[k]| \leq \frac{C}{|k|^{p+1+\varepsilon}} \quad \text{ for some }
C, p, \varepsilon >0. \nonumber
\end{align}
\end{fact}

\subsection{Approximation analysis and its definitions}
\label{sec:sysdef}

For $i = 1, \ldots, M$, bandlimited approximations of $f_i(t)$ with Fourier
coefficient~$a_i[k]$ are defined as
\begin{align}
\label{eq:Fourierapp} \widehat{f}_i(t) = \sum_{k=-L}^L b_i[k] \exp(j 2\pi kt)
\quad \mbox{ for } t \in [0,1],
\end{align}
where $b_i[-L], \ldots, b_i[L]$ will be a function of $a_i[k], k \in \zZ$ chosen
later according to specified criterion. Here, $L$ the \textit{bandwidth} of the
approximate signals $\widehat{f}_i(t)$.

Let the \textit{sum signal analytic} of interest be $s(t) = f_1(t)+ \ldots +
f_M(t)$. From $s(t)$, for example, $\max_t s(t)$ or $\mbox{avg}\{s(t)\}$ can be
obtained. The corresponding approximation obtained by the IoT cloud is
$\widehat{s}(t) = \widehat{f}_1(t) + \ldots + \widehat{f}_M(t)$.  Let $E :=
d(s(t), \widehat{s}(t))$ be the \textit{approximation error} according to some
distance measure.  For a given bandwidth $L$, the overpredictive distributed
signal analytics problem (see Section~\ref{sec:intro}) is:
\begin{align}
\min_{\widehat{f}_i(t)} E := d(s(t), \widehat{s}(t)) \quad \text{ subject to }
\widehat{s}(t)  \geq s(t).
\end{align}
%
%
%
For a signal $f(t)$, the approximation $\widehat{f}_{\env}(t)$ such that
$\widehat{f}_{\env}(t) \geq f(t); \forall t \in [0,1]$ is called an
\emph{envelope approximation}~\cite{kumar2017}. 
%
%
%
%

\subsection{Distance measures of interest}

For overpredictive approximation, we will construct envelope of the signal
$f(t)$ recorded by an IoT device~\cite{kumar2017}.  While making envelope
approximation $f_{\env}(t)$ of a signal $f(t)$, a distance function is needed to
capture the proximity of $f_{\env}(t)$ with $f(t)$~\cite{kumar2017}. In this
work, the ${\cal L}^1, {\cal L}^2$ and ${\cal L}^\infty$ distance measures will
be used.
Using the Fourier representations of $f_{\env}(t)$ (akin
to~\eqref{eq:Fourierapp}) and $f(t)$, and the envelope property $f_{\env}(t)
\geq f(t)$ it can be observed that 
\begin{align}
\|f_{\env}-f\|_1 & = b[0] -a[0],\label{eq:L_1} \\
\|f_{\env}- f \|_2^2 = & \sum_{|k| \leq L} |b[k] -a[k]|^2 + \sum_{|k|>L}
|a[k]|^2 \label{eq:L_2} \\
\mbox{and, } \|f_{\env} - f\|_\infty & \leq \sum_{|k| \leq L} |b[k] -a[k]| +
\sum_{|k|>L} |a[k]|. \label{eq:L_infi}
\end{align}    
In the above equation, the ${\cal L}^{\infty}$ distance is upper-bounded using
an $\ell_1$ error between $b[k]$ and $a[k]$ in the sequence space.  The
approximation error in signal analytics $E = d(s(t), \widehat{s}(t))$
corresponding to the ${\cal L}^1, {\cal L}^2, {\cal L}^\infty$ distance measures
will be denoted by $\SA_1, \SA_2, \SA_\infty$, respectively.

\subsection{Bandwidth and approximation error tradeoff}
\label{sec:pro_statement}
The fundamental tradeoffs between the bandwidth parameter $L$ and the
approximation error $E$ for ${\cal L}^p, p = 1, 2, \infty$ distance metrics will
be presented,
with the constraint that $\widehat{s}(t) \geq s(t)$.

%
%

%
%

\section{Overpredictive signal analytics}
\label{sec:Algo}
Recall the smart city applications described in Fig.~\ref{fig:SCApp}, with $M$
IoT devices reporting the approximate signals observed to the IoT cloud.
The signal analytics problem of interest are
\begin{align}
\min \SA_q := \int_{0}^1 \left| \widehat{s}(t) - s(t) \right|^q \mbox{d}t \mbox{
subject to } \widehat{s}(t) \geq s(t) \label{eq:sa1sa2}
\end{align}
for $q = 1, 2$, and
\begin{align}
\min \SA_\infty = \| \widehat{s} - s \|_\infty \mbox{ subject to }
\widehat{s}(t) \geq s(t). \label{eq:sainf}
\end{align}
where the above minimizations are over 
$\widehat{f}_1(t), \ldots, \widehat{f}_M(t)$.

\subsection{Algorithm for overpredictive signal analytics in IoT}

It is assumed that each IoT device works in a distributed manner. To ensure
$\widehat{s}(t) \geq s(t)$, we propose that each IoT device can perform envelope
approximation of its observed signal~\cite{kumar2017}.  The following steps 
are proposed for obtaining a bandwidth-$L$ approximation $\widehat{s}(t)$ of
$s(t)$:
\begin{enumerate}
\item Each device $i$ records its individual signal $f_i(t)$, calculates its
envelope $\widehat{f}_{i,\env}(t)$, and communicates its $(2L+1)$ Fourier
coefficients to the IoT cloud.
\item Using Fourier coefficients from each IoT device, the 
cloud calculates $\widehat{s}_{\env}(t) = \widehat{f}_{1,\env}(t) + \ldots +
\widehat{f}_{M,\env}(t)$.
%
%
\end{enumerate}
%

Signal envelope calculation in Step~1 above is outlined next. For ${\cal L}^1$
distance (see~\eqref{eq:L_1}), it will be calculated as~\cite{kumar2017}
\begin{align}
\mbox{minimize } & b[0] - a[0] \nonumber \\
\mbox{subject to } & \vec{b}^T \Phi(t) \geq f(t) 
\end{align}
where $\Phi(t) = [\exp(-2\pi L t), \ldots, \exp(2 \pi Lt)]^T$ and $\vec{b} =
(b[-L], \ldots, b[L])^T$ are the Fourier series coefficients of the envelope
approximation. The above linear program with linear constraints is solvable
efficiently~\cite{kumar2017}. For ${\cal L}^2$ and ${\cal L}^\infty$, the cost
function $b[0] - a[0]$ is replaced by those in~\eqref{eq:L_2}
and~\eqref{eq:L_infi}.

As $L$ is increased, the envelopes $\widehat{f}_{i, \env}(t)$ become more
proximal to their target $f_i(t)$. It is expected that $\SA_1$, $\SA_2$, and
$\SA_\infty$ will decrease as $L$ increases.  However, analyzing the dependence
of $\SA_q, q = 1, 2, \infty$ versus $L$ is difficult. Accordingly, na\"{i}ve
envelope approximation~\cite{maheshwari2015} will be used to analyze fundamental
bounds on their tradeoff. Since each IoT device approximates the signal in a
distributed manner, the error in $\SA_q, q = 1, 2, \infty $ will increase with
the number of IoT devices in the network as discussed in the next
section.

%
%
%

\subsection{The na\"{i}ve envelope approximation}
\label{subsec:DC_boost}
First consider IoT device $1$ in isolation.  Let $f_{1,\proj}(t)$ be the
orthogonal projection of $f_1(t)$ on the span of
$\exp{(j2\pi kt)}$ for $|k| \leq L$.  Then $f_{1,\proj}(t) = \sum_{|k|\leq L}
a_1[k] \exp({j2\pi kt}).$ The na\"{i}ve envelope approximation scheme is as
follows~\cite{maheshwari2015}:
\begin{align} 
f_{1,\env}(t) = f_{1,\proj}(t) + C_0, \label{eq:naive_app}
\end{align}
where $C_0 = \|f_1 - f_{1, \proj}\|_\infty$. Using the triangle inequality, 
\begin{align}
C_0  \leq \sum_{|k|>L} |a_1[k]| \leq \sum_{|k|>L} \frac{C}{|k|^p}, \qquad p>1.
\label{eq:C0bound}
\end{align}
For $p > 1$~\cite[Sec.~2.2]{bhatia1993}
%
%
we can show that, $C_0 = O\left(\frac{1}{L^{p-1}}\right)$.



\section{Na{i}ve envelope approximation analysis}
\label{sec:MainRes}
\begin{figure*}[!htb]
\centering
\includegraphics[scale=0.88]{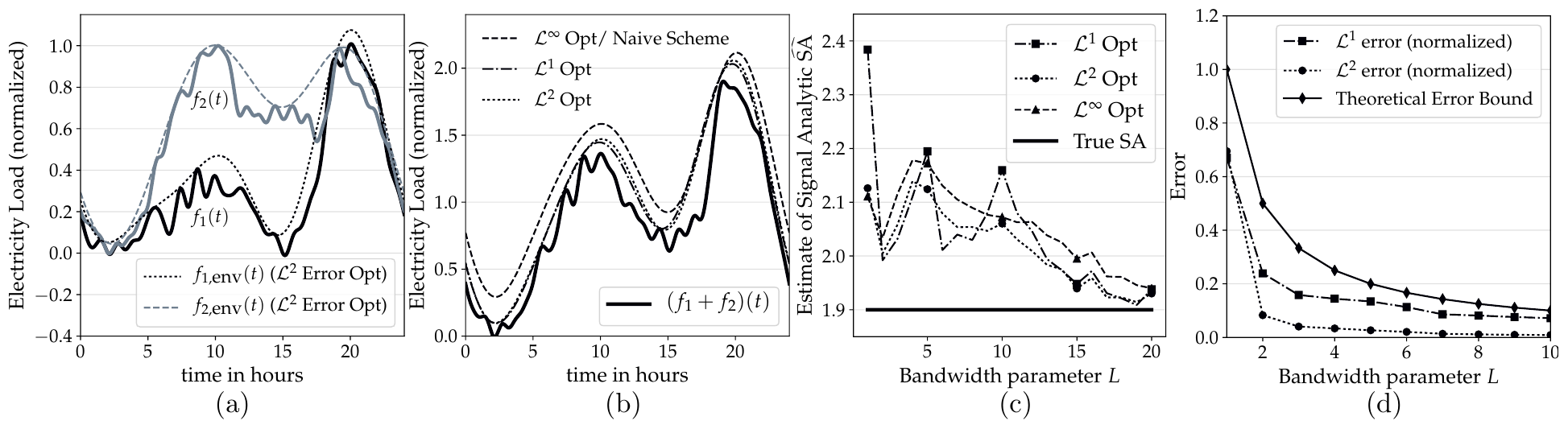} 
  \caption{\label{fig:ForEdit} (a) Distributed envelope approximation applied on
  electricity load datasets collected from Eastern and Western Region power
  grids in India \cite{posoco2016} for $L=3$ approximation coefficients (b) Sum
  envelope recovered at the IoT cloud using $\mathcal{L}^q\;\; q =1,2,\infty$
  error optimization (c) Convergence of $\widehat{\SA}$ to $\SA$ with increasing
  available bandwidth $L$ at IoT device (d) Error versus Bandwidth plots for
  $\mathcal{L}^1$ and $\mathcal{L}^2$ error metric compared to the theoretical
  upperbound with $C_2(q=1,2) = 0.5$ (see Theorem~\ref{th:main}).}
\end{figure*}
\begin{figure*}[!htb]
\centering
\includegraphics[scale=0.88]{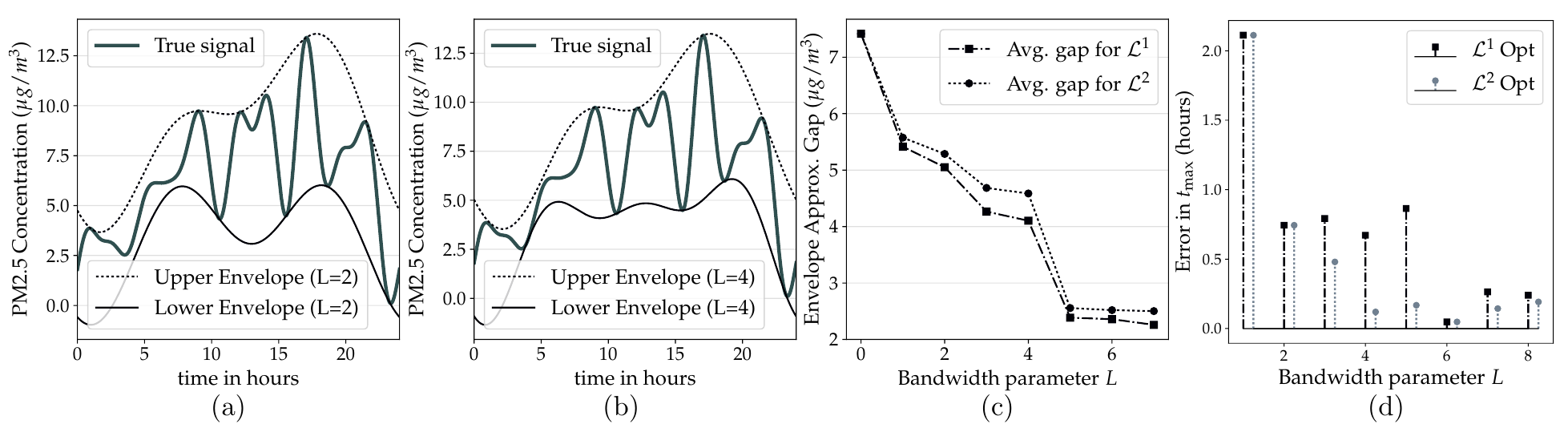} 
	\caption{\label{fig:ForEdit2}Dataset : PM2.5 (Speciation) concentration
	in Denver from US-EPA database \cite{us-epa17}(a) Lower and upper
	envelopes for $L=2$ with $\mathcal{L}^2$ optimization (b) Lower and
	upper envelopes for $L=4$ with $\mathcal{L}^2$ optimization (c) The
	variation of average gap between upper and lower envelopes with
	increasing values of $L$ (d) Error in time estimate of peak pollution
	concentration, $t_{\mbox{\scriptsize max}}$ for various value of $L$.}
\end{figure*}
The main result is proved in this section.
\begin{theorem}
%
%
\label{th:main}
In~\eqref{eq:sa1sa2} and~\eqref{eq:sainf}, let $\SA_q$ be the optimal distance
and $\SA'_q$ be the distance corresponding to the na\"{i}ve approximation
in~\eqref{eq:naive_app}.  There exist signals $f_1(t), \ldots, f_M(t)$ such
that for $q = 1, 2, \infty$
\begin{align}
C_1 (q) \leq \frac{\SA'_q}{\SA_q} \leq C_2 (q) \mbox{ for } 0 <
C_1(q) < C_2(q) < \infty. \nonumber
\end{align}
%
%
%
\end{theorem}

\textit{Proof}:  Using \eqref{eq:C0bound} and the triangle inequality on sum of
signals and their envelopes, we determine bounds on $\SA_q$. They are tabulated
in Table~\ref{tab:UB_LB}, where the result follows for $q = 2, \infty$ by taking
ratios of $\SA_q$ and $\SA'_q$. The steps are omitted due to space constraints.
This result shows that the naive approximation is \emph{order optimal}
for the $\mathcal{L}^2$ and $\mathcal{L}^\infty$ errors, if the IoT
device signals are $p$-times differentiable.

The optimality of the na\"{i}ve envelopes for the $\SA_1$ distance holds 
for a signal class with the following properties: (i) the Fourier coefficients
$a_1[k] \geq 0$, and (ii) $f_1(t)$ is real and even, that is $a_1[k] = a_1[-k]$.
>From these symmetry assumptions, it follows that $b_1[k] = b_1[-k]$.  Restricted 
to this signal class, the $\SA_1$ envelope approximation is re-stated as:
\begin{align}
\argmin_{b_1[k], |k| \leq L} b_1[0] &, \quad \text{subject to}  \nonumber \\
b_1[0] + 2 \sum_{ 1 \leq |k| \leq L}  b_1[k] \cos(2\pi k t)  & \geq \sum_{|k|>L}
a_1[k] e^{j 2\pi k t} \label{eq:envapp_L1}
\end{align}
The above optimization can be shown to result in $b_{1,\text{\scriptsize opt}}
[0] = \sum_{|k|>L} a_1[k]$ if $a_1[k] = 0$ for $|k|<L$. In this case $a_1[0]=0$.
And, for this signal $\SA_1' = C_0 = \|f_1 - f_{1,\proj}\|_\infty \leq
\sum_{|k|>L} a_1[k]$ as all the $a_1[k]$ Fourier coefficients are positive.
Since $\SA_1' \geq \SA_1$, as naive method will be suboptimal, so the two are
equal.

It is noted that $\SA_1$ and $\SA'_1$ in Table~\ref{tab:UB_LB} are comparable.
The presented result is for 1 IoT devices.  For $M$ IoT devices, $\SA_1$ and 
$\SA_\infty$ as well as $\SA'_1$ and $\SA'_\infty$ scale linearly with $M$. And 
$\SA_2$ as well as $\SA'_2$ scale quadratically with $M$. Their ratios remain 
the same as in $M= 1$.

\begin{table}[!htb]
\centering
\caption{\label{tab:UB_LB} Bounds on the approximation errors}
  \bgroup \def\arraystretch{1.15} \begin{tabular}{lcc}
%
\toprule
$\mathcal{L}^q$ error metric & $\SA_q$  & $\SA'_q$     \\[1.2ex] \midrule
  $q = 1$  &  $b_1[0] - a_1[0]$  &  $\begin{aligned}\sum_{|k|>L}
  |a_1[k]|\end{aligned}$ 
\\[3ex] 
  $q = 2$ &  $\begin{aligned} \frac{2}{2p-1}\frac{1}{(L+1)^{2p-1}}\end{aligned}$
    &  $\begin{aligned} \frac{2}{2 p-1}\frac{1}{L^{2p-1}} \end{aligned}$ \\[3ex] 
      $q = \infty$ &
      $\begin{aligned}\frac{2}{p-1}\frac{1}{(L+1)^{p-1}}\end{aligned}$  &
	$\begin{aligned} \frac{2}{p-1}\frac{1}{L^{p-1}}\end{aligned}$  \\[3ex]
	  \bottomrule
\end{tabular} \egroup
\end{table}
%

%
%
%
%
%
%
%
%
%

%

\section{Numerical simulations}
\label{sec:Numsim}
\textit{Dataset Description}: Time series data of electricity
loads (in KWhr) from Eastern and Western region grids of India are considered
\cite{posoco2016} (see Fig.~\ref{fig:ForEdit}). The time samples available at 30
minute intervals are interpolated using low pass projection on Fourier basis
with 25 coefficients. For brevity of simulations a normalized amplitude scale is
used.
We also consider the pollution dataset from US EPA\footnote{United States
Environmental Protection Agency} \cite{us-epa17}, consisting of the
time-series variation of PM2.5 concentration in Denver. The hourly sampled data,
for a duration of 1 day, is smoothened using lowpass filtering using 10 Fourier
coefficients (see Fig.~\ref{fig:ForEdit2}).

\textit{Simulation setup and analysis}: Simulations for the approximation error
and bandwidth tradeoffs are presented for grid load data in
Fig.~\ref{fig:ForEdit}~\cite{posoco2016}. The time series plots for individual
load variation are shown in Fig.\ref{fig:ForEdit} (a). The sum signal envelope is
reconstructed  at the IoT Cloud with $(2L+1) = 7$  Fourier coefficients (see
Fig.~\ref{fig:ForEdit}(b)).  The convergence of the estimate $\widehat{s}(t)$ to
$s(t)$ with increasing $L$ is studied in Fig.~\ref{fig:ForEdit}(c).
It is observed that the maxima of the sum signal is eventually
tracked under each of the cost functions. The oscillations in the error plot (see
Fig.~\ref{fig:ForEdit} (c)) are the artifacts of the distributed and nonlinear
nature of the approximation algorithm. At any IoT device, the maximum
(nonlinear) of the signal is tracked with uniformly decreasing error. However,
since each device independently (ie in a distributed manner) reports the
approximations to the IoT cloud, the errors do not die down uniformly, but with
occasional rise. Fig.~\ref{fig:ForEdit}(d) captures the bandwidth-error
tradeoff for the three distance measures considered in the paper. The
performance of the na\"{i}ve approximation scheme is seen to coincide with that of
the $\mathcal{L}^\infty$ error optimization.  This is attributed to the
relaxation of the $\mathcal{L}^\infty$ cost in terms of the coefficients as
discussed in \eqref{eq:L_infi}.

We analyze the gap between the upper and lower signal envelopes
using the pollution dataset \cite{us-epa17} (Fig.~\ref{fig:ForEdit2}).  We
observe that the average gap between the envelopes decays with increasing value
of $L$. It is to be noted that the lower envelope is negative in certain time
intervals\footnote{This indicates the limitation of Fourier basis representation
in capturing the signal for a generic signal class. Alternate basis
representations such as polynomials or wavelets suitable for the signal class
will be considered as future extensions.}. In Fig.~\ref{fig:ForEdit2} (d) we
analyze the error in time estimate of the peak pollutant concentration,
$t_{\mbox{\scriptsize max}}$.  The estimates approach the true value of signal
analytic (ie $t_{\mbox{\scriptsize max}}$) eventually with growing bandwidth
parameter $L$.

\begin{figure}[!hbt]
	\centering
	\includegraphics[scale=0.5]{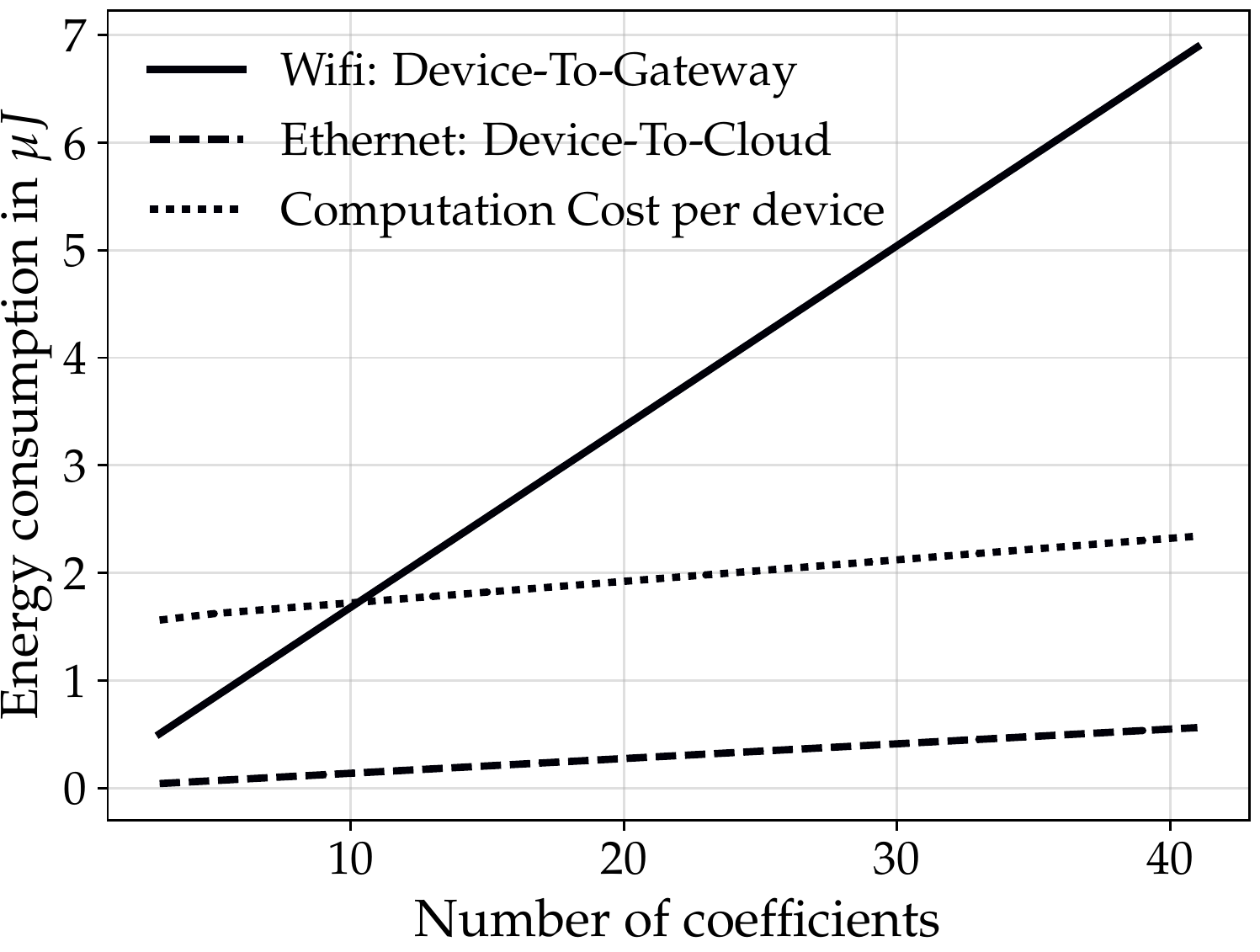}
	\caption{\label{fig:Energy_Cons}Energy expenditure at IoT device for
  communication and computing.  Simulation parameters - $E_b = 0.4279 \;nJ/$bit
  for Ethernet,  $5.25 \;nJ/$bit for WiFi, bits per coefficient $B_c=32$ bits  and
  $100$ GFLOPS/W processor.}
\end{figure}

\textit{Energy consumption in IoT Device and IoT Cloud} :By using IoT
energy consumption data, a comparison of  different IoT energy requirements are
shown in Fig.~\ref{fig:Energy_Cons}~\cite{ISOC-IOT-2015}.  For a
communication  mode with $E_b$ energy/bit requirement and $B_c$
bits/coefficient, the proposed envelope scheme requires $(2L+1)E_b B_c$  energy
units per device. The computing energy requirement (see
Fig.~\ref{fig:ForEdit}(d)) is determined by counting the number of
floating point operations (FLOP) in MATLAB to execute the optimization
algorithm at the device. The conversion of FLOP to equivalent Joules is
performed with respect to a 100 GFLOP/sec/W processor (see TI C667x processor
\cite{ti-dspC667x}). It is observed that the communication and computation
energy expenditure is observed to linear increase with the number of
approximation coefficients.

\section{Conclusions}
Distributed nonlinear signal analytics was introduced for the first time. It was
observed that in applications such as energy distribution, signal observed by
IoT devices should be subjected to envelope approximation. An algorithm for
over-predictive nonlinear signal analytics in the IoT cloud was developed in
this work, by using the envelope approximation technique.  The fundamental
tradeoff between approximation error in signal analytics ($\SA_q, q = 1, 2,
\infty$) and the bandwidth parameter $L$ was established. It was observed that
this tradeoff depends on the smoothness of signals at the IoT devices.
Simulation results were presented for load data collected from two power grids.
Envelope approximation schemes using polynomial and wavelet basis are proposed as a
future extensions.

\bibliographystyle{IEEEbib}
\bibliography{refs}

\end{document}